\documentclass{elsart}

\usepackage{graphicx}

\usepackage{amssymb}

\begin{document}

\begin{frontmatter}



\title{Tracking in Antiproton Annihilation Experiments}


\author[lnf]{Olaf N. Hartmann}

\address[lnf]{Laboratori Nazionali di Frascati dell'INFN \\ 
Via E. Fermi,40, 00044 Frascati (RM), Italy}

\begin{abstract}
A major ingredient of the planned new accelerator complex FAIR, 
to be constructed at the GSI, Darmstadt, Germany,
is the availability of
antiproton beams with high quality and intensity. Among the experiments
which will make use of this opportunity is PANDA, a dedicated
experiment to study antiproton annihilations on nucleons and nuclei. This
article gives an overview on the foreseen techniques to perform charged
particle tracking in the high rate environment of this experiment.
\end{abstract}

\end{frontmatter}

\section{Introduction}
\label{intro}

With the future FAIR \cite{fair} facility at the GSI \cite{gsi} in Darmstadt, 
Germany, beams of
antiprotons will become available. After production using a primary proton
beam, collecting and cooling, the antiprotons will be delivered to a
dedicated synchrotron/storage ring (named HESR), where they will be stored,
accelerated/decelerated to the desired energy, cooled and used for
experiments with an internal target (p, $A$). The momentum range of the
$\overline{\mbox{p}}$ beam will be between 1.5 and 15 GeV/c, i.e. the maximum
$\overline{\mbox{p}}$p center-of-mass energy will be 5.5 GeV. The design 
luminosity is
2$\cdot$10$^{32}$ cm$^{-2}$s$^{-1}$, and a momentum spread of $\delta p/p
\approx 10^{-4}$ is envisaged. With a somewhat relaxed demand in luminosity,
$\delta p/p\approx 10^{-5}$ should be achievable. 
These parameters translate into an
annihilation rate of about 10$^7$ s$^{-1}$ = 10 MHz, thus despite having a 
moderate multiplicity per event the experiment has to deal with high rates.

\section{Experimental Evironment}
\label{expsetup}

\begin{figure}[h]
\includegraphics[width=\textwidth]{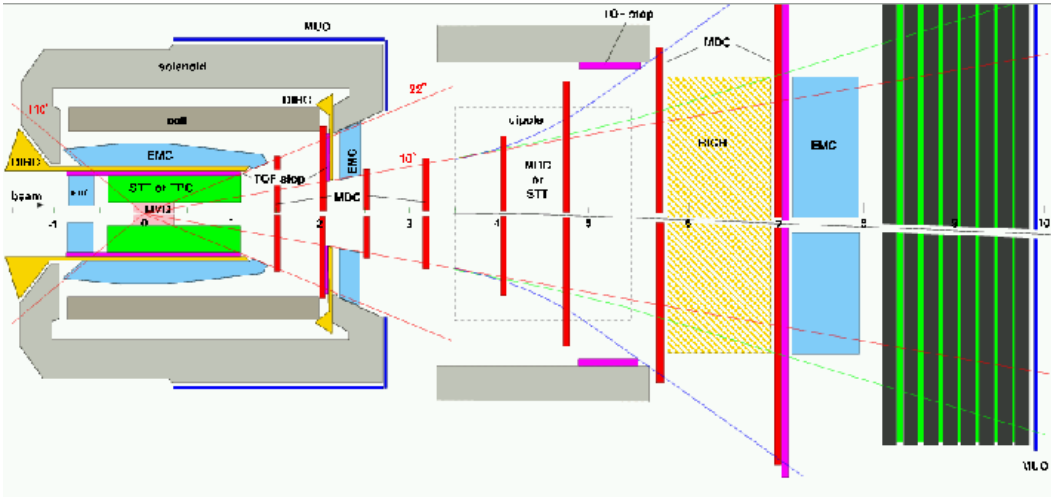}
\caption{\label{pandatopview} Top view of the PANDA setup.}
\end{figure}

The antiproton annihilations occuring on the internal target are to be studied
by the PANDA \cite{panda} detector. PANDA is a fixed target experiment which
consists of a target spectrometer surrounding the interaction point and a
forward spectrometer to measure down to 0$^{\circ}$ in polar angle. Figure
\ref{pandatopview} shows the schematic view (from the top) of the whole setup. The figure is taken from \cite{tpr}. Further details of the experimental setup 
can be found in \cite{tpr}.\\

The requirements to the charged particle tracking systems are: 
a solid angle coverage of nearly 4$\pi$, the capability to resolve multiple tracks, good spatial resolution for secondary vertices, high momentum resolution $O(1\%)$
in a range from MeV/c to several GeV/c, minimal material budget,
high rate capability (the occupancy reaches values of
$\approx 3\cdot 10^4$ cm$^{-2}$s$^{-1}$ in the forward region) and 
robustness against ageing effects. Finally it operates in a
magnetic field. As indicated in Fig. \ref{pandatopview}, the experiment
applies a solenoid field (2 T) for the target spectrometer and a dipole field
(2 Tm) for the forward part.\\
The components of the tracking system are:
\begin{itemize}
\item a microvertex detector (MVD) close to the interaction point (IP), 
using silicon pixel/silicon strip detectors
\item a central tracker surrounding the MVD, made from straw tubes, or, as
an alternative under study, using a time projection chamber
\item drift chambers in the target and the forward spectrometer for the
particles emitted under small polar angles
\end{itemize}
In the following, this article focuses on the straw tube central tracker;
more information and details about the other tracking devices can be found 
in \cite{tpr}.

\section{The central tracker}
\label{stt}

\begin{figure}[h]
\includegraphics[width=.5\textwidth]{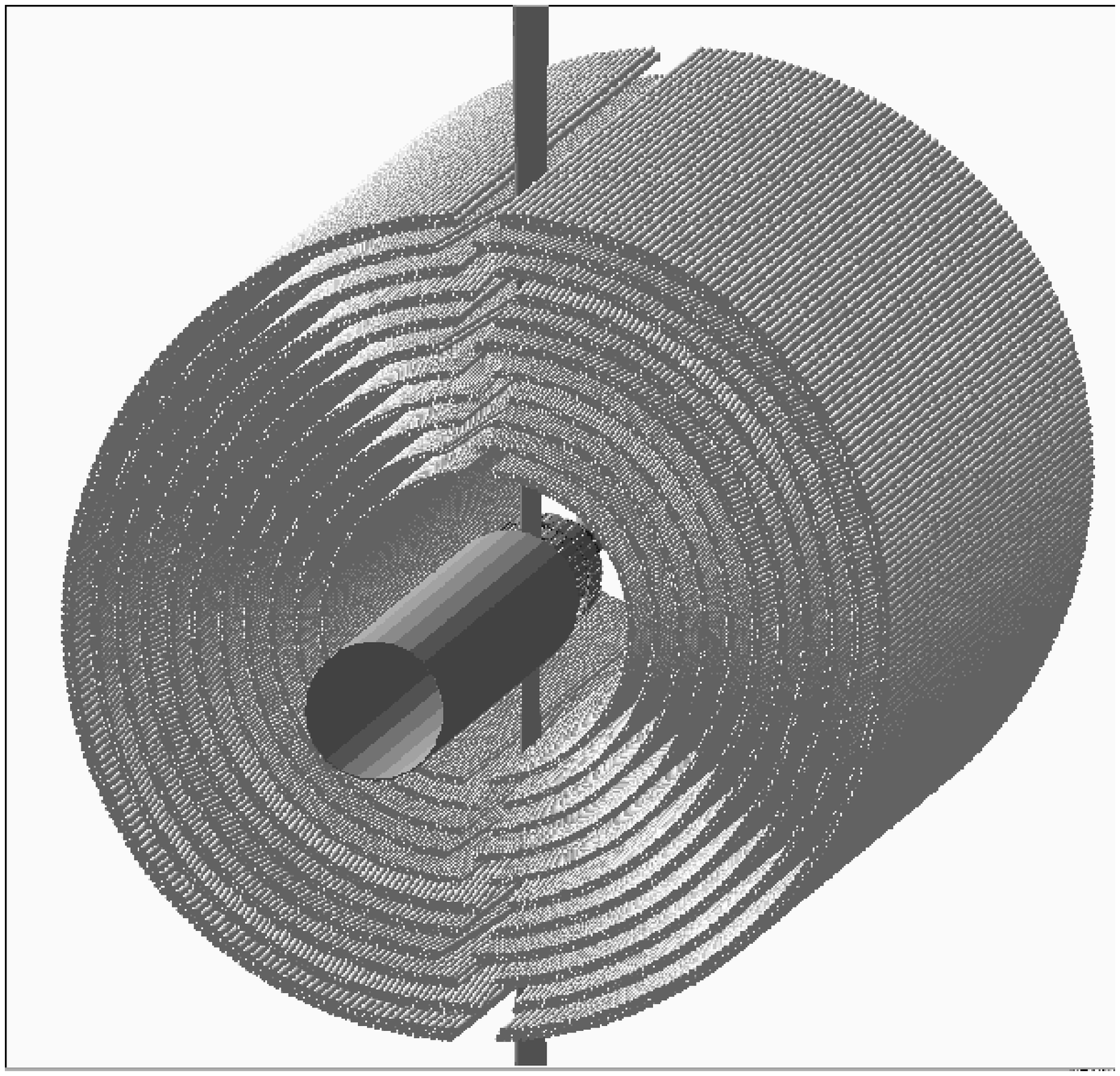}
\includegraphics[width=.5\textwidth]{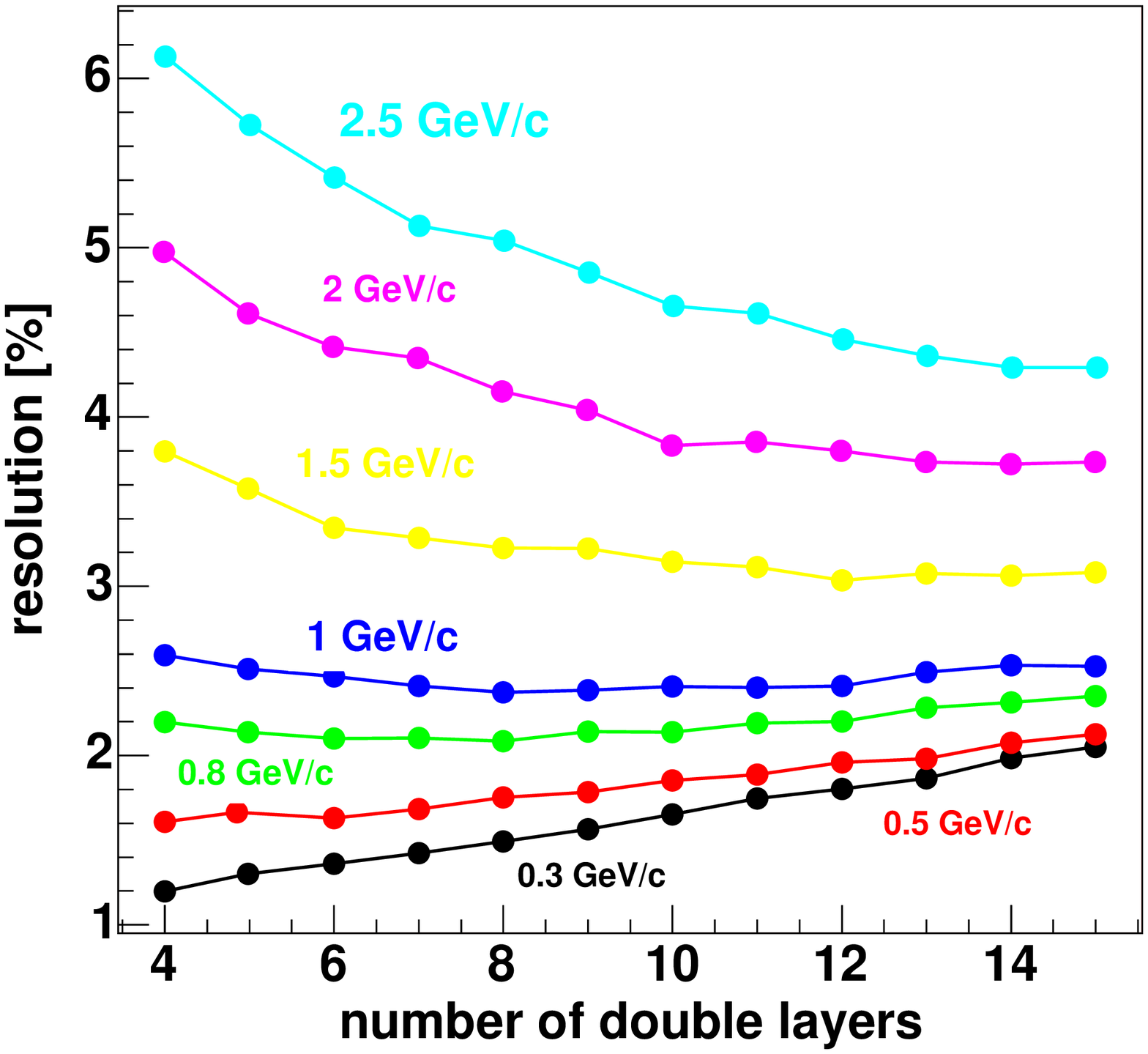}
\caption{\label{stt_11} Left: The central tracker made from straw tubes. The
support structures are not shown. Right: Relative comparison of the
resolution as a function of the number of double layers for various particle 
momenta.}
\end{figure}

The straw tube tracker (STT) is a cylindrical object consisting out of
several double layers of straw tubes oriented along the beam axis. It should
cover the radial space from 15 to 42 cm with respect to the beam axis. In
the left part of Fig. \ref{stt_11}, a possible arrangement using 11 double
layers is shown; the beam pipe in the center and the perpendicular target
pipe are shown as well.\\
The tubes have a diameter of 6 mm for the inner layers and 8 mm for the
outer layers; 20 $\mu$m $\varnothing$ W-Re is used for the anode wires, and a mylar/kapton 
film with Al coating for the tubes. The
counting gas has been chosen to be a mixture of 90\% Ar and 10\% CO$_2$,
at normal or slight overpressure.
In such a configuration, one tube amounts to 0.05\% X/X$_0$ material budget,
and the full device to approximately 1\% (excluding cables and support structures at the downstream/upstream end).

\begin{figure}
\includegraphics[width=.5\textwidth]{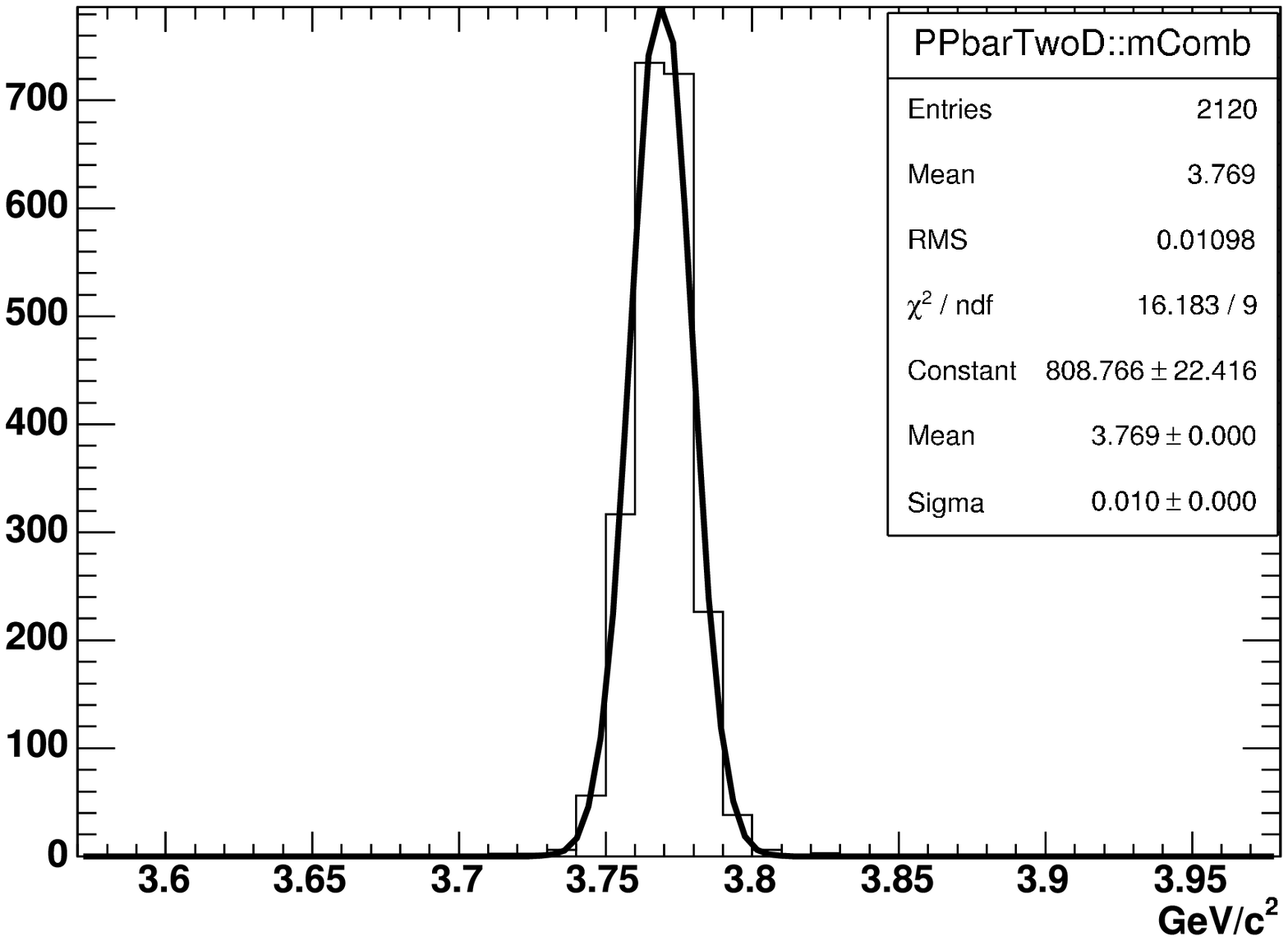}
\includegraphics[angle=90,width=.5\textwidth]{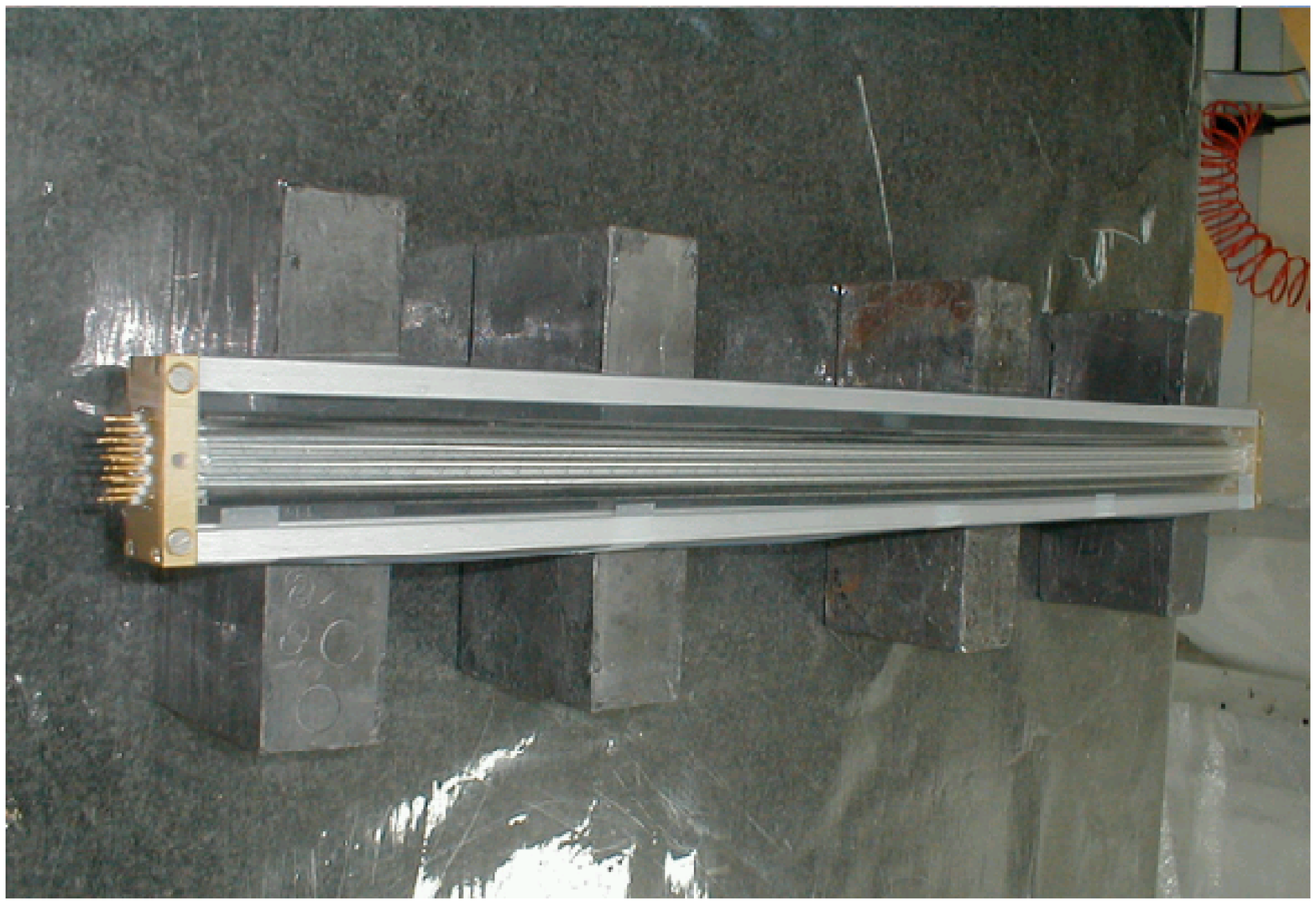}
\caption{\label{ddbar} Left: Simulation of a physics channel using STT.
Right: Photograph of a straw tube prototype prepared at the LNF.}
\end{figure}

The basic arrangement in double layers is needed to resolve the left-right 
ambiguity of
the passage of a charged particle with respect to the wire. In the right
part of Fig. \ref{stt_11} a relative comparison of the resolution as a
function of the number of double layers is shown for different momenta of the
traversing particles, which for this simulation are $\mu^-$ emitted under an
angle of 45$^{\circ}$ from the nominal interaction point. The resolution has been extracted
from the width of a gaussian fit to the appropriate 1/$p$ distributions
\cite{lia}. As one can see, with a number of 11($\pm 1$) double layers the
requirements for low and high momentum particles can be satisfied, adding more double layers does not improve the overall resolution. Together with an
additional information from the vertex detector, the momentum resolution reaches a level of 1.5 to 2\% over the whole quoted momentum range.\\[1ex]
In the left part of Fig. \ref{ddbar} the result of a physics channel using a
full scale Monte Carlo \cite{tpr} with the STT for particle tracking is shown as an example: the $\Psi(3770)$
resonance, formed in a $\overline{\mbox{p}}$p annihilation, decays into a
D$\overline{\mbox{D}}$ pair. The channel
D$^{\pm}\rightarrow$K$^{\mp}\pi^{\pm}\pi^{\pm}$ is reconstructed. 
Here, the invariant mass
distribution is plotted; the $\Psi(3770)$
can be reconstructed with a resolution of
$\sigma_\Psi\approx10$ MeV/c$^2$ and a reconstruction efficiency of
$\approx$4\%.

Concerning the gas mixture, He and Ar based mixtures have been studied in
a simulation \cite{tpr,andrei}. Finally, the above mentioned Ar:CO$_2$ 9:1
mixture has been chosen since it has a high number of primary ionisations
per track length, and thus gives a better efficiency and spatial resolution. 
The efficiency, measured with a 10 mm $\varnothing$ straw \cite{wintz}
is $\approx$99\% over the major part of the tube; very
close to the tube wall (last 0.25 mm) it drops to $\approx$75\% due to the 
low number of primary ionisations along the short track piece.
The drift time is smaller than 100 ns and thus matches the expected interaction rates.
The spatial resolution of such a detector is $< 150\mu$m in $r\varphi$.
Since the tubes have the readout only on one side, the
$z$ coordinate instead could be obtained by using stereo tubes, i.e. rotating
several double layers by a small so-called $skew$ angle of 2-3$^{\circ}$ to
resolve the ambiguities. With this method in principle a $z$ resolution of 
several mm can be achieved. Under investigation is the possibility to connect 
two neighbouring tubes, thus allowing a two-side readout, and to obtain the $z$
coordinate by charge division. As discussed in \cite{andrei}, one could expect a resolution of 7 to 15 mm. Nevertheless, at present
only a resolution in the order of $>2$ cm could be reached in tests with a
radioactive source, further improvement is needed to make this method 
competitive with the skewed tubes.\\
The geometry and the mechanical
mounting of the STT is fairly complicated due to the presence of beam- and 
target pipe and the lack of
overall space (see Fig. \ref{stt_11}, right, and Fig. \ref{pandatopview}) 
inside the coil of the
solenoid. These questions are currently under study. Finally, the right part of Fig.
\ref{ddbar} shows a photograph of a straw tube prototype constructed at the 
INFN Frascati laboratories to be used for further R\&D.

\section{Acknowledgments}
\label{ack}
I acknowledge the support of the European Community-Research
Infrastructure Activity under the
FP6 ``Structuring the European Research Area'' programme (HadronPhysics,
contract number RII3-CT-2004-506078).




\begin{thebibliography}{00}

\bibitem{fair} {\bf F}acility for {\bf A}ntiproton and {\bf I}on {\bf R}esearch, http://www.gsi.de/fair
\bibitem{gsi} {\bf G}esellschaft f\"ur {\bf S}chwer{\bf i}onenforschung mbH, http://www.gsi.de/
\bibitem{panda} $\overline{\mbox{\bf P}}$ {\bf An}nihilation at {\bf Da}rmstadt,
http://www.gsi.de/panda
\bibitem{tpr} Strong Interaction Studies with Antiprotons, Technical
Progress Report for PANDA, Darmstadt, 01/2005
\bibitem{lia} L. Lavezzi, private communication
\bibitem{wintz} P. Wintz et al., AIP conference proceedings, 698 (2004) 789
\bibitem{andrei} A. Sokolov, Dissertation, University of Giessen, 2005







\end{thebibliography}
\end{document}